\documentclass[aps,amsmath,amssymb,prl,twocolumn,showpacs]{revtex4}
\usepackage{bm}
\usepackage{epsfig}

\newcommand{\C}{{\mathcal{C}}}
\newcommand{\bnabla}{\mbox{\boldmath$\nabla$}}
\newcommand{\rav}[1]{\left<#1\right>_{\textit{d}}}
\newcommand{\tav}[1]{\left<#1\right>_{\textit{th}}}
\newcommand{\paper}{Letter }

\begin{document}

\title{Displacement Profile of  Charge Density Waves and
Domain Walls at Critical Depinning}
\author{Andreas Glatz}
\author{Thomas Nattermann}
\affiliation{Institut f\"ur Theoretische Physik, Universit\"at zu
K\"oln, Z\"ulpicher Str. 77, 50937 K\"oln, Germany}

\date{\today}

\begin{abstract}
The influence of a strong surface potential on the critical
depinning of an elastic system driven in a random medium is
considered. If the surface potential prevents depinning completely
the elastic system shows a parabolic displacement profile. Its
curvature $\mathcal{C}$ exhibits at zero temperature a pronounced
rhombic hysteresis curve of width $2f_c$ with the bulk depinning
threshold $f_c$. The hysteresis disappears at non-zero
temperatures if the driving force is changed adiabatically. If the
surface depins by the applied force or thermal creep,
$\mathcal{C}$ is reduced with increasing velocity. The results
apply, e.g., to driven magnetic domain walls, flux-line lattices
and charge-density waves.
\end{abstract}

\pacs{75.60.-d, 74.60.Ge}

\maketitle

The driven viscous motion of an interface in a medium with random
pinning forces is one of the paradigms of condensed matter physics
\cite{Fisher98,Kadar98}.  This problem arises, e.g., in the domain
wall motion of magnetically or structurally ordered systems with
impurities \cite{review.magnetic.wall} or when an interface
between two immiscible fluids is pushed through a porous medium
\cite{Rubio89}. Closely related problems are the motion of other
elastic systems like a vortex line in an impure superconductor
\cite{Blatter94}, of a dislocation line in a solid
\cite{Ioffe.Vinokur86} or driven charge density waves (CDWs)
\cite{Thorne96}. For a constant external driving force this
problem has been considered close to the zero temperature critical
depinning threshold \cite{Natter+92,NarFish93,Ertas+94,Chauve+01}
and in the creep region \cite{Ioffe.Vinokur86,Nattermann86}.

It was a tacit assumption of  these investigations that the motion
of the elastic system is not hindered by effects from surfaces or
internal grain boundaries. Surface barriers are however known to
be relevant in all  cases mentioned above. In superconductors they
prevent the penetration of new flux lines into the probe
\cite{Zeldov}. In CDWs normal electrons have to be converted into
those condensed in the CDW by a phase--slip mechanism which is
essentially a nucleation process \cite{Monceau98,Rama92}. The
motion of domain walls may be hindered by a variation of the width
of the sample such that position of minimal width are preferred
etc.

It is  the aim of the present \paper to consider the effect of a
strong surface pinning potential in addition to the weak bulk
random pinning. It turns out  that, starting from a flat
interface, at $T=0$ and increasing the driving force $f$ to
$f>f_c$ the average displacement profile shows a parabolic profile
with a mean curvature $\C(f)=(f-f_c)/\Gamma$. In more general
situations $\C(f,t)$ exhibits a pronounced hysteretic behavior. At
non--zero temperatures $\C(f,t)$ increases with time and reaches
asymptotically its value of the pure system $f/\Gamma$. We further
determine the reduction of the  curvature in the case that the
surface is depinned due to a sufficiently large driving force or
due to thermally activated processes at the surface. The latter
mimic also phase slip processes in CDWs.


{\it Model and zero temperature critical depinning}.--- We focus
on a simple realization of the problem. The equation of motion of
a D-dimensional field $\varphi({\bf x},t)$ describing an interface
profile in the case of domain walls or a phase profile in the case
of CDWs is given by $\frac{1}{\gamma}\frac{\partial
\varphi}{\partial t}=-\frac{\delta{\mathcal H}}{\delta\varphi}$,
where ${\mathcal H}$ denotes the Hamiltonian of the system:
   \begin{equation}
   {\mathcal H}=\int d^Dx\left\{\frac{\Gamma}{2}(\mathbf{\nabla}\varphi)^2-
   f\cdot\varphi+V({\bf x},\varphi)\right\}\,.\label{eq:H}
   \end{equation}
$\gamma$ and $\Gamma$ denote the mobility and the stiffness
constant of the elastic object, respectively, and $f$ is the
driving force which is assumed to change only adiabatically. The
potential includes a random force and a surface contribution
   \begin{equation}
   V({\bf x},\varphi)=-\int\limits_0^{\varphi}d\varphi^{\prime}
   g({\bf x},\varphi^{\prime})[1-\rho({\bf x})]+\frac{\Gamma}{a^2} V_s(\varphi)\rho({\bf x})\,.
   \label{eq:V_R}
   \end{equation}
The random force $-g({\bf x},\varphi)$ is assumed to be Gaussian
distributed with $\rav{g}=0$ and $\rav{g({\bf x},\varphi)g({\bf
x}^{\prime},\varphi^{\prime})}=\delta^{(D)}({\bf x}-{\bf
x}^{\prime})\Delta_0(\varphi-\varphi^{\prime})$ where
$\rav{\ldots\phantom{l}}$ denotes the random average. For domain
walls $\Delta_0(\varphi)=\Delta_0(-\varphi)$ is an analytical
monotonically decreasing function of $\varphi$ which decays to
zero over a finite distance $l$. For CDWs $g\propto
\sin(\varphi-\alpha({\bf x}))$ with a random phase $\alpha({\bf
x})\in[0,2\pi[$ and therefore $\Delta_0(\varphi)$ is periodic with
$\Delta_0(\varphi)=\Delta_0(\varphi+2\pi\mathbb{Z})$.

The surface potential $V_s(\varphi)$ is assumed to act only in the
vicinity ($a\ll L$) of  $x_1=0$ and $x_1=L$, e.g., $\rho({\bf
x})=e^{-x_1/a}+e^{(x_1-L)/a}$, and favors the values of
$\varphi(0,{\bf x}_{\perp})$ and $\varphi(L,{\bf x}_{\perp})$ at
$2\pi \mathbb{Z}$. The details of the interaction between the
elastic system and the surface depend on the specific system under
consideration. We will restrict ourselves here to a periodic
surface potential which has applications in type-II
superconductors and may also serve as a first step for the
treatment of conversion phenomena in CDWs.

The case $V_s\equiv 0$ was considered previously in
\cite{Natter+92,NarFish93,Ertas+94,Chauve+01}. It was shown that
at zero temperature the system undergoes a depinning transition at
a critical value $f_c$. For $f>f_c$ the velocity
$v=\big<\dot\varphi\big>$ increases as $v\sim(f-f_c)^{\beta}$ with
the critical exponent $\beta$ calculated in an expansion in
$D=4-\epsilon$ dimensions. The average displacement profile is
macroscopically flat. At non--zero temperatures the depinning
transition is smeared out and goes over into a creep motion for
$f\ll f_c$ \cite{Ioffe.Vinokur86}.

In this \paper we will consider the opposite case where a strong
surface potential $V_s$ slows down or prevents completely the
motion of the elastic object. Then the displacement profile
becomes parabolic with a history-dependent curvature $\C(f,t)$ .
The steady state solution for the average phase is given by
   \begin{equation}
   \varphi_0\equiv\big<\varphi\big>=vt+\frac{\C_s(f)}{2}(L-x_1)x_1\,.
   \label{eq:parabola}
   \end{equation}

where $\C_s(f)=\C_s(f,t\to\infty)$ is  the saturation value of the curvature.


{\it Infinite surface barriers}.--- We begin with the case
$V_s\to\infty$, where the depinning transition is depressed.  To
determine $\C(f,t)$ we first apply perturbation theory. Using the
decomposition $\varphi({\bf x})=\varphi_0({\bf x})+\varphi_1({\bf
x})$ with $\rav{\varphi_1({\bf x})}=0$ in the equation of motion
and expanding $g({\bf x},\varphi_0+\varphi_1)$ to linear order in
$\varphi_1$ we get after averaging over the disorder
\begin{equation}
 \frac{1}{\gamma}\dot\varphi_0=-\Gamma \C(t)+f+\rav{g_{\varphi}({\bf
 x},\varphi_0({\bf x},t))\varphi_1({\bf x},t)}\,,\label{eq:pert0}
\end{equation}
where $g_{\varphi}({\bf
x},\varphi)=\frac{\partial}{\partial\varphi}g({\bf x},\varphi)$.
Calculating $\varphi_1$ also  to first order of $g$ we get from
(\ref{eq:parabola}) and (\ref{eq:pert0}) $\mathcal
{C}_s=\mathcal{C}_0=f/\Gamma$ since $\Delta^{\prime}_0(0)=0$,
i.e., there seems to be no influence of the disorder. Here, the
situation is completely analogous to that at the conventional
depinning transition \cite{Natter+92}. However, as we know from
critical depinning, this is the situation below the Larkin scale
$L_p$.

Next we discuss   renormalized perturbation theory starting from a
situation where ${\mathcal C}_s=0$. As long as $f\le f_c$, the
elastic object is pinned and  boundary pinning does not matter,
hence $\mathcal{C}(f,t)=0$. At $f=f_c$
the elastic object is in the same critical state as at the
depinning transition. Therefore we can use the results of the
previous renormalization group calculation in this case. As a
result $\gamma$ and $\Delta_0(z)$ are replaced there by the
renormalized, momentum dependent quantities
\begin{subequations}
\begin{eqnarray}
&\gamma(p)\simeq\gamma(pL_P)^{-2+z}\,,&\label{eq:gp}\\
&\Delta_{p}(\varphi)\approx K_D^{-1}(\Gamma l/L_P^{\zeta})^2
p^{4-D-2\zeta}\Delta^{\ast}\left(\varphi(p
L_P)^{\zeta}/l)\right).& \label{eq:replacement}
\end{eqnarray}
\end{subequations}
$\zeta$ denotes the roughness exponent which was
calculated for domain walls to order $\epsilon=4-D$ in
\cite{Natter+92,NarFish93} and recently to ${\mathcal
O}(\epsilon^2)$ \cite{Chauve+01}. For CDWs $\zeta=0$
\cite{NarFish93}. The most important feature of
$\Delta_p(\varphi)$ is that $\Delta^{\ast}(\varphi)$ has a
cusp--like singularity at the origin. The renormalized equation
for $\mathcal{C}(f,t\to\infty)$ is given by
\begin{eqnarray}
&&\Gamma
\mathcal{C}(f,t)=f+f_p\omega_p\int\limits_0^{\infty}dt^{\prime}\,\int\limits_0^1d\tilde
p\,\times \label{eq:a_renorm}\\
&&\times\tilde p^{1+z-\zeta}e^{-\omega_p\tilde p^z
t^{\prime}}\Delta^{\ast\prime}\left([-\mathcal{C}(t)+\mathcal{C}(t-t^{\prime})]\frac{x_1}{2}(x_1-L)\right)\,,\nonumber
\end{eqnarray}
where $\tilde{\bf p}={\bf p}L_p$, $\omega_p=\gamma f_p/l$ and
$f_p=l\Gamma L_p^{-2}$. After having increased $f$ adiabatically
to a fixed value slightly larger than $f_c$, $\mathcal{C}(f,t)$
saturates for $t\rightarrow \infty$ and hence the difference
$\mathcal{C}(t)-\mathcal{C}(t-t^{\prime})$ vanishes. As a result
the argument of ${\Delta^{\ast}}^{\prime}$ also vanishes and the
right hand side of (\ref{eq:a_renorm})  becomes independent of ${
x_1}$.
Since $\mathcal{C}(t)>\mathcal{C}(t-t^{\prime})$ the argument of
$\Delta^{\ast\prime}$ approaches zero from positive values. Thus
we get for the saturation value $\C_s(f)$
\begin{equation}
\mathcal{C}_s(f)=\frac{f-f_c}{\Gamma
}=\frac{l}{L_p^2}\frac{f-f_c}{f_p},\quad\quad
f_c=\frac{f_p}{2-\zeta}\Delta^{\ast\prime}(0^+)\,.\label{eq:a_s}
\end{equation}
One can understand this result in the following way: Using the
decomposition $\varphi=\varphi_0+\varphi_1$ in the asymptotic
region, where $\mathcal{C}(t)$ saturates, the equation of motion
can be written as
\begin{equation}
\frac{1}{\gamma}\dot\varphi_1=\Gamma\bnabla^2\varphi_1+f-\Gamma
\mathcal{C}+g_1({\bf x},\varphi_1)\label{eq:motion1}
\end{equation}
where $g_1({\bf x},\varphi_1)=g({\bf x},\varphi_0({\bf
x})+\varphi_1({\bf x}))$. $g_1({\bf x},\varphi)$ and $g({\bf
x},\varphi)$ have the same statistical properties. According to
(\ref{eq:motion1}) the force acting on the field $\varphi_1$ is
now reduced by the curvature force $-\Gamma\mathcal{C}$. The
depinning of the $\varphi_1$-field seems hence to occur at
$f\nearrow\tilde f_c=f_c+\Gamma \mathcal{C}_s$. However, since the
boundary conditions fix $\varphi_1(0)=\varphi_1(L)=0$ and hence
$\langle\dot\varphi_1\rangle=0$ for all values of $f$, the system
{\it is always at its depinning transition}, which implies
(\ref{eq:a_s}). Starting from some $f< f_c$ and $\mathcal{C}_s=0$,
$\mathcal{C}_s$ will stay at this value until $f$ reaches $f_c$.
For $f>f_c$, $\mathcal{C}_s$ obeys (\ref{eq:a_s}). The same
argument can be used for negative forces $f<0$. Then we find for
$f<-f_c$: $\Gamma \mathcal{C}_s(f)=f+f_c=-(|f|-f_c)$ since
$\Delta^{\ast\prime}(0^-)=-\Delta^{\ast\prime}(0^+)$.


A {\it scaling argument} supports the validity of eq.
(\ref{eq:a_s}) to all orders in $g$: Close to the $V_s=0$
depinning transition the correlation length $\xi$ diverges as
$\xi\approx L_p\left(({f-f_c})/{f_c}\right)^{-\nu}$. For
$L^{\prime}<\xi$ the {\it roughness}  -- the mean square
displacement of a piece of linear size $L^{\prime}$ of the elastic
object -- scales as $w^2(L^{\prime})\approx
l^2\big(L^{\prime}/L_p\big)^{2\zeta}$
\cite{Natter+92,NarFish93,Ertas+94,Chauve+01}. If we choose the
system size $L\approx\xi$ we expect that the roughness scales as
the height of the parabolic $\varphi-$profile on the same scale,
\textsc{} $w(\xi)\approx \mathcal{C}\xi^2$, which is indeed
fulfilled if we use the scaling law $\nu={1}/({2-\zeta})$
   \begin{equation}
   \mathcal{C}=\frac{w(\xi)}{\xi^2}=
   \frac{l}{L_p^2}\left(\frac{f-f_c}{f_c}\right)^{\nu(2-\zeta)}
   \approx \frac{f-f_c}{\Gamma}\,.
   \label{eq:a}
   \end{equation}


{\it Hysteresis}.--- Next we consider the case that we increase
$f$ adiabatically from $f\lesssim f_c$ to a value
$f_{\text{max}}$, where $\mathcal{C}(f,t)$ reaches
$\mathcal{C}_{\text{max}}$, and then {\it decrease} $f$ again. In
this case $\mathcal{C}(f,t)<\mathcal{C}(f,t-t^{\prime})$ and hence
the argument of $\Delta^{*\prime}$ becomes negative. Instead of
(\ref{eq:a_s}) we get from (\ref{eq:a_renorm})
\begin{equation}
\Gamma \mathcal{C}_{\text{max}}\equiv f_{\text{max}}-f_c=f+f_c\,.
\end{equation}
The effective force acting on the elastic object is now given by
$f-\Gamma \mathcal{C}_{\text{max}}$. Further decreasing $f$, there
is no change of $\mathcal{C}(f,t)$ until the effective force
reaches the threshold $-f_c=f-\Gamma
\mathcal{C}_{\text{max}}=f-(f_{\text{max}}-f_c)$. According to the
last relation this happens at $f=\tilde
f_{\text{max}}=f_{\text{max}}-2f_c$. Analogous arguments can be
used for reversing the fields from $\dot f<0$ to $\dot f>0$. Thus
$\mathcal{C}_s$ undergoes a {\it hysteresis} which consists of the
two parallel segments given by $\mathcal{C}_s\Gamma=(f\mp f_c)$
and two horizontal segments determined by
$\mathcal{C}_{\text{max}}=(f_{\text{max}}-f_c)/\Gamma$ and
$\mathcal{C}_{\text{min}}=(f_{\text{min}}+f_c)/\Gamma$,
respectively. Indeed, similar hysteresis effects of the strain
have been observed in CDWs \cite{Thorne01}.

\begin{figure}[htb]
\includegraphics[width=0.95\linewidth]{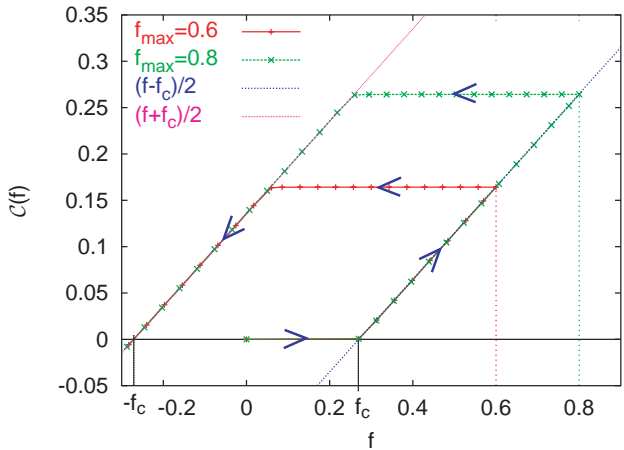}
\caption{Hysteresis of $\mathcal{C}(f)$ at $T=0$ for an
one-dimensional interface. The driving force is first increased to
$f_{\text{max}}=0.6$ or $f_{\text{max}}=0.8$, respectively, and
then decreased to $-f_c\approx -0.27$. The arrows show the
direction of the hysteresis. The numerical simulation was done for
an interface with length $L=1000$ and averaged over $300$ disorder
configurations. }\label{fig1}
\end{figure}

These findings are fully supported by numerical simulations, as
shown in fig. \ref{fig1}. For integrating the equation of motion,
the ${\bf x}$--coordinate is discretized with a lattice constant
$\alpha$ and the simulation time is measured in units of a time
$\tau_0$ (the dimensionless lattice laplacian for $D=1$ is given
by $\bnabla^2\varphi_i=\varphi_{i+1}+\varphi_{i-1}-2\varphi_i$,
with lattice sites $i=0,\ldots,L$). $\alpha$ and $\tau_0$ are
chosen such that $\frac{\tau_0\gamma\Gamma}{\alpha^2}=1$ and the
dimensionless stochastic forces $\tau_0\gamma g({\bf
x},\varphi)\in [-1/2,1/2]$ (the dimensionless driving force is
$\tau_0\gamma f$).

\begin{figure}[htb]
\includegraphics[width=0.8\linewidth]{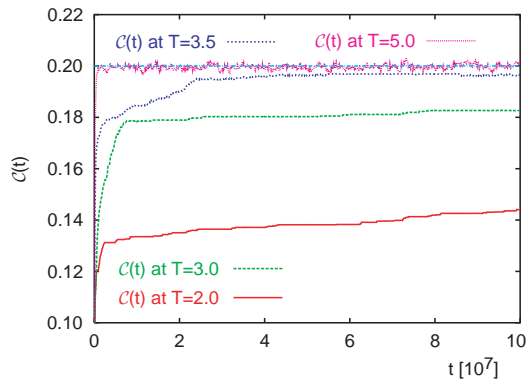}
\caption{Simulation--time resolved coefficient $\mathcal{C}(t)$
for a driving force of $f=0.2$ at various temperatures. The
simulation was done for a system of length $L=1000$ and for one
disorder configuration of CDW--type for each temperature (see
text).}\label{fig2}
\end{figure}

{\it Curvature at finite temperature}.--- Next we want to consider
the problem of finite temperatures. Changing $f$ only
adiabatically we may use equilibrium statistical mechanics. It is
convenient to go over to the field $\tilde \varphi({\bf x}) =
\varphi({\bf x}) + \frac{f}{2\Gamma}x_1(x_1-L)$. The  Hamiltonian
rewritten in $\tilde\varphi$ has the same statistical properties
as the initial one (\ref{eq:H}), since $V_R({\bf
x},\varphi)=-\int^{\varphi}d\varphi^{\prime} g({\bf
x},\varphi^{\prime})$ is a {\it random} function of both
arguments. This can most easily seen by using the replica method
\cite{review.magnetic.wall}. The disorder averaged free enthalpy
follows from the {\it replica Hamiltonian}
\begin{equation}
{\mathcal H}_n=\frac{\Gamma}{2}\sum_{a,b=1}^n \int_{\bf
x}\left\{(\bnabla \tilde \varphi_a)^2\delta_{a,b}-
\frac{\Gamma}{T} R(\tilde \varphi_a-\tilde
\varphi_b)\right\}\,,\label{eq:replica.Hamiltonian}
\end{equation}
with $\rav{V_R({\bf x},\varphi)V_R({\bf x}^{\prime},
\varphi^{\prime})}=\delta^{(D)}({\bf x}-{\bf
x}^{\prime})R(\varphi-\varphi^{\prime})$ Apparently, the replica
Hamiltonian is the same as that following from (\ref{eq:H}). It is
worth to mention that this is true only if the random potential
$V_R(\bf x,\varphi)$ is strictly uncorrelated in $\bf
x$-direction. The application of surface barriers implies
therefore $\mathcal{C}={f}/{\Gamma}$ and
$\tav{\langle(\tilde\varphi({\bf x})-\tilde\varphi({\bf
x}^{\prime}))^2\rangle_d}^{1/2}\simeq
l\left({L}/{L_p}\right)^{\tilde \zeta}$ where $\tilde\zeta$
denotes the equilibrium roughness exponent corresponding to
Hamiltonian (\ref{eq:replica.Hamiltonian}). Thus the displacement
profile is the same as in the pure case. For non--adiabatic
changes of $f$, traces of the $T=0$ hysteresis are expected to be
seen at non--zero temperatures (cf. fig. \ref{fig2}).

The numerical solution of equation of motion with thermal noise at
finite temperatures and $V_s=\infty$ is in agreement with these
analytical considerations. Fig. \ref{fig2} shows the coefficient
$\C(t)$ as it approaches its saturation value $\C_s=f/(2\Gamma)$
with time.  Strictly speaking, we are not in a steady state until
$\C(t)$ has reached its saturation value and hence the phase
profile deviates slightly from the parabolic shape. In fig.
\ref{fig2}, $\C(t)$ is the least square fit to the profile. Note,
that for {\it low} temperatures ($T<5.0$ in the simulation, where
$T$ is the dimensionless variance of the thermal noise) this
approach is very slow, noticeable by the occurring steps,
triggered by avalanches, even at large times. For high $T$
($T=5.0$), one sees that $\C(t)$ fluctuates around the saturation
value due to thermal noise. Therefore the $T=0$ hysteresis of $\C$
vanishes at finite temperatures.


{\it Critical Depinning}.--- So far the surface potential was
assumed to fix the value of the displacement field $\varphi$ at
the surfaces $x_1=0$ and $x_1=L$. We will now assume that the
surface potential is reduced (or temperature is raised from zero)
such that a macroscopic motion of the elastic object is possible.
To determine the mutual interaction between the bulk and the
surface we have to consider the effective equation of motion of
the surface. Denoting $\varphi(0, {\bf x}_\bot)=\varphi_s({\bf
x}_\bot)$  the effective equation of motion of the surface field
can be written as ($a$ is assumed to be of the order of the
lattice spacing)
\begin{equation}
\frac{1}{\gamma}\dot\varphi_{s}=\Gamma \mathcal{C}\frac{L}{2a}+
\Gamma\bnabla_{\bot}^2\varphi_{s}+f -\frac{\Gamma}{a^2}
V^{\prime}_s(\varphi_s).\label{eq:motion2}
\end{equation}
An analogous equation can be written for $\varphi(L, {\bf
x}_\bot)$. In (\ref{eq:motion2}) we have replaced the force
resulting from the displacement in the bulk by the corresponding
average force. In the steady state $\bnabla_{\bot}^2\varphi_{s}=0$
and eq. (\ref{eq:motion2}) has a depinning threshold $f_{s,c}\gg
f_c$ determined by
\begin{equation}\label{threshold}
\Gamma \mathcal{C}(f_{s,c})\frac{L}{2a}+f_{s,c}
-\frac{\Gamma}{a^2} \text{max} V^{\prime}_s(\varphi)=0
\end{equation}
For $f>f_{s,c}\gtrsim f_c$ the macroscopic velocity is given by
the steady state solution $v=\dot\varphi_{s}$ which follows from
integrating (\ref{eq:motion2}) with
$\bnabla_{\bot}^2\varphi_{s}=0$. The corresponding solution
\begin{equation}\label{velocity.surface}
  v(t)= v_p \Phi\left(\frac{\Gamma\mathcal{C}\frac{La}{2}+fa^2}{\Gamma V_{s,max}^{\prime}},t\right)
\end{equation}
depends of course on the specific form of the surface potential,
$v_p=\gamma f_p$. Eq.(\ref{velocity.surface}) has to be combined
with the effective equation for the bulk ($f>f_c$)
\cite{Natter+92}
\begin{equation}\label{velocity.bulk}
  \left(\frac{v(t)}{v_p}\right)^{1/\beta}=
  \frac{f-f_c}{f_p}-\mathcal{C}(t)\frac{L_p^2}{l}
\end{equation}
which follows from (\ref{eq:gp}) and (\ref{eq:a_s}). Eqs.
(\ref{velocity.surface}) and (\ref{velocity.bulk}) determine both
the velocity and the curvature $\mathcal{C}$ as a function of the
driving force.  If we increase $f$ from $f=0$ with $\C=0$,
$\mathcal{C}$ remains zero until we reach $f_c$. For
$f_c<f<f_{s,c}$, $\mathcal{C}$ obeys (\ref{eq:a_s}). At $f_{s,c}$
the elastic object is depinned and curvature is reduced with
increasing velocity, according to (\ref{velocity.bulk}). If the
surface potential is periodic, also $v(t)$ will be periodic and
the bulk depinning transition is slightly smeared out
\cite{Glatz03}. We will assume that this effect is weak. In
principal it can be avoided by adding some randomness to the
surface potential.


{\it Nucleation and Creep}.--- At finite but low temperatures the
surface field may exhibit a creep motion even if $f\ll f_{s,c}$.
Creep proceeds via the formation of  droplets at the surfaces
$x_1=0$ and $x_1=L$, inside which $\varphi$ is changed by $ 2\pi$,
respectively, with respect to the bulk value of $\varphi$. The
droplet  consist of a cylindrical piece (the cylinder axis is
perpendicular to $x_1=0$) in the surface layer of height $a$  and
radius $R$ and an attached semi-sphere with the same radius. The
width of the droplet wall confining the cylinder is of the order
$a^{\prime}=a/\sqrt{V_s^{\prime\prime}}$.

Keeping only the leading order terms we get for the energy of the
droplet $E_{\text{dp}}(R)=2\Gamma
R^{D-2}\left\{\sqrt{V_s^{\prime\prime}}+\ln\frac{R}{a^{\prime}}-\pi\C
RL-fR^2/\Gamma \right\}$. The critical droplet size $R_C\ll L$
follows as $R_c\approx \sqrt{V_s^{\prime\prime}}/(\mathcal{C}L)$
or $R_c=(\mathcal{C}L)^{-1}$ in $D=3$ or $D=2$, respectively. In
deriving $E_{\text{dp}}$, we have neglected the contribution from
the disorder which is correct as long as $R_c<L_p$, i.e., $L\gg
L_p$. The nucleation rate of droplets and hence the creep velocity
is given by
\begin{equation}\label{creep.velocity}
  \frac{v_{D=3}}{v_p}=A\exp\left(-B\frac{V_s^{\prime\prime}\Gamma}{\mathcal{C}LT}\right)
  ,\,\,
  \frac{v_{D=2}}{v_p}=A^{\prime}\left(\frac{\mathcal{C}La^{\prime}}{\sqrt{V_s^{\prime\prime}}}\right)^{B^{\prime}\Gamma/T}
\end{equation}
which replaces (\ref{velocity.surface}) in the case  $f\ll
f_{s,c}$, $T>0$ (cf. \cite{Rama92}). The present treatment is too
crude to give the coefficients $A,A^{\prime},B,B^{\prime}$. Again,
(\ref{creep.velocity}) has to be considered together with
(\ref{velocity.bulk}) to determine $\mathcal{C}$ and $v$.
 In  CDWs, where $\varphi$ can be multi-valued, nucleation processes
also occur deep in the bulk \cite{Rama92}. The droplet energy then
does not contain a term $\sim f$,  leaving the relations
(\ref{creep.velocity}) essentially unchanged.

To conclude we have shown that surface pinning of impure elastic
systems lead to an onset of curvature $\mathcal{C}$ only above a
threshold value $f_c$ of the random force. In general,
$\mathcal{C}$ exhibits a pronounced hysteresis. The curvature is
reduced  above the surface depinning  transition or at finite
temperatures when nucleation processes at the surface allow for
creep motion.

We thank B. Rosenow, S. Brazovskii, and in particular T. Emig for
fruitful discussions. The authors acknowledge support by
Sonderforschungsbereich 608.

\end{document}